\documentclass[superscriptaddress,reprint,aps,prl]{revtex4-1}
\usepackage{amssymb, amsmath}
\usepackage{graphicx}
\usepackage{dcolumn}
\usepackage{bm}
\usepackage{float}
\usepackage{xcolor}
\usepackage[normalem]{ulem}

\begin{document}

\title{ $\mu$SR study of triangular Ising antiferromagnet ErMgGaO$_{4}$}


\author{Y. Cai}
\author{C. Lygouras}
\author{G. Thomas}
\author{M. N. Wilson}
\author{J. Beare}
\address{Department of Physics and Astronomy, McMaster University, Hamilton, Ontario L8S 4M1, Canada}
\author{D.R. Yahne}
\address{Department of Physics, Colorado State University, Fort Collins, CO 80523, USA}
 \author{K. Ross}
 \address{Department of Physics, Colorado State University, Fort Collins, Colorado 80523, USA}
 \address{Canadian Institute for Advanced Research, Toronto, Ontario M5G 1M1, Canada}
\author{Z. Gong}
\author{Y. J. Uemura}
\address{Department of Physics, Columbia University, New York, New York 10027, USA}
\author{H.A. Dabkowska}
\address{Brockhouse Institute for Materials Research, McMaster University, Hamilton, Ontario, Canada L8S 4M1}
\author{G.M. Luke}
\address{Department of Physics and Astronomy, McMaster University, Hamilton, Ontario L8S 4M1, Canada}
\address{Canadian Institute for Advanced Research, Toronto, Ontario M5G 1M1, Canada}
\address{TRIUMF, Vancouver, British Columbia. V6T 2A3, Canada}
\begin{abstract}
We present a detailed magnetic study of the triangular antiferromagnet ErMgGaO$_4$. A point charge calculation under the single ion approximation reveals a crystal field ground state doublet with a strong Ising-like behavior of the Er$^{3+}$ moment along the c axis. Magnetic susceptibility and specific heat measurements indicate no presence of magnetic transitions above 0.5~K and no evidence of residual entropy as temperature approaching zero. Zero field (ZF) $\mu$SR measurements shows no sign of static uniform or random field and longitudinal field (LF) $\mu$SR measurements exhibit persistent spin fluctuations down to our lowest temperature of 25~mK. Our results provide evidence of a quantum spin liquid state in the triangular antiferromagnet ErMgGaO$_4$.
\end{abstract}

\maketitle
A quantum spin liquid (QSL) is a state of matter in which spins are highly entangled and do not show magnetic order down to zero temperature \cite{Leon}. QSL's are of great current interest both from a fundamental physics point of view and for possible applications in quantum computation \cite{08_Nayak}. Geometrically frustrated magnets (where competing magnetic interactions cannot be simultaneously satisfied) are excellent candidate materials for QSL behavior since magnetic order is suppressed in them by the frustration \cite{06_greedan, 10_gardner, 91_reimers}. Such frustration frequently arises from antiferromagnetically coupled spins located on triangle-based lattices (stacked triangular, kagome, pyrochlore), and can lead to a highly degenerate ground state without magnetic order. The previously studied quasi-two dimensional triangular layered material YbMgGaO$_4$ (with YbFe$_2$O$_4$-type structure) has been attracting considerable interest as a potential quantum spin liquid candidate \cite{Li, Li2, Paddison, Shen, Yuesheng, Zhu, Cevallos2}. 

YbMgGaO$_4$ has a Curie-Weiss temperature of $\sim$ -4~K but shows no sign of long-range order down to 30~mK \cite{Li, Li2, Paddison, Shen}. Its magnetic specific heat in zero field shows a broad hump at 2.4~K instead of sharp $\lambda$-type peak which would be expected for a well-defined second order phase transition. The magnetic excitation spectra appears as a broad continuum in inelastic neutron scattering measurements, which has been taken as an evidence for a QSL state \cite{Paddison, Shen}. Particularly, the anisotropic exchange interactions between rare earth ions are found to be playing a crucial role in stabilizing such spin liquid ground state \cite{Li, Li2, Paddison, Yuesheng, Shen}, although an alternative explanantion in terms of the random Ga/Mg site mixing has also been proposed \cite{Zhu}. Such exchange interactions associated with spin orbit coupling strongly depends on rare earth ions. It has also been found, particularly, in rare earth pyrochlores ($R_2 B_2$O$_7$ with $R$~$=$~rare earth, $B$~$=$~non-magnetic cation), that the interplay of exchange couplings, dipolar interactions, and single ion anisotropy leads to spin glasses \cite{91_reimers, 14_silverstein}, spin liquids \cite{17_wen, 19_gao, 19_gaudet}, spin ices \cite{01_snyder}, order-by-disorder \cite{03_champion, 14_ross}, magnetic moment fragmentation \cite{16_benton, 16_petit}, and conventional long-range magnetic ordering \cite{Stewart}. These observations indicating different rare earth ions can result in much different ground states, motivating the search for RMgGaO$_4$ with different magnetic rare earth elements.

In this paper, we report our study of the stacked triangular compound ErMgGaO$_4$. We have successfully synthesized single crystals of ErMgGaO$_4$ using the floating zone technique. On the basis of point charge calculations, we find a crystal field ground state doublet for Er$^{3+}$ with strong Ising anisotropy along local [001] axes. We have investigated the collective magnetic properties of this system with magnetic susceptibility, heat capacity and muon spin rotation and relaxation. We find no magnetic transition down to our lowest temperature of 25~mK using ZF-$\mu$SR and no residual entropy which suggests there will not be any further magnetic transitions down to zero temperature. LF-$\mu$SR measurements indicate the presence of persistent spin fluctuations down to our lowest temperature 25~mK, a feature in common with some other high frustrated magnetic systems including SrCr$_{9p}$Ga$_{12-9p}$O$_{19}$ \cite{SCGO} and Tb$_2$Ti$_2$O$_7$ \cite{TbTiO7}. Collectively, our observations indicates that ErMgGaO$_4$ is an Ising-like quantum spin liquid candidate.

Crystals of ErMgGaO$_4$ were prepared in an optical floating zone image furnace at McMaster University \cite{Yipeng}. Stoichiometric mixture powders of pre-annealed Er$_2$O$_3$, Ga$_2$O$_3$ and MgO were pre-reacted at 1200$^{\circ}$C in air for 12 hours. The powder was later made into rods (6 cm in length and 8 mm in diameter) through hydrostatic pressure at 60~MPa for 15 minutes and these rods were then sintered at 1450$^{\circ}$C in oxygen for 56 hours with an intermediate grinding and reformation. Both pre-reacted powder and sintered rods were examined by powder X-ray diffraction. No pure phase of ErMgGaO$_4$ was found in either diffraction pattern and both contained magnetic Er$_3$Ga$_5$O$_{12}$ and non-magnetic MgO, indicating the desired phase was not formed during the sintering process. 
\begin{figure}
\includegraphics[width=\columnwidth]{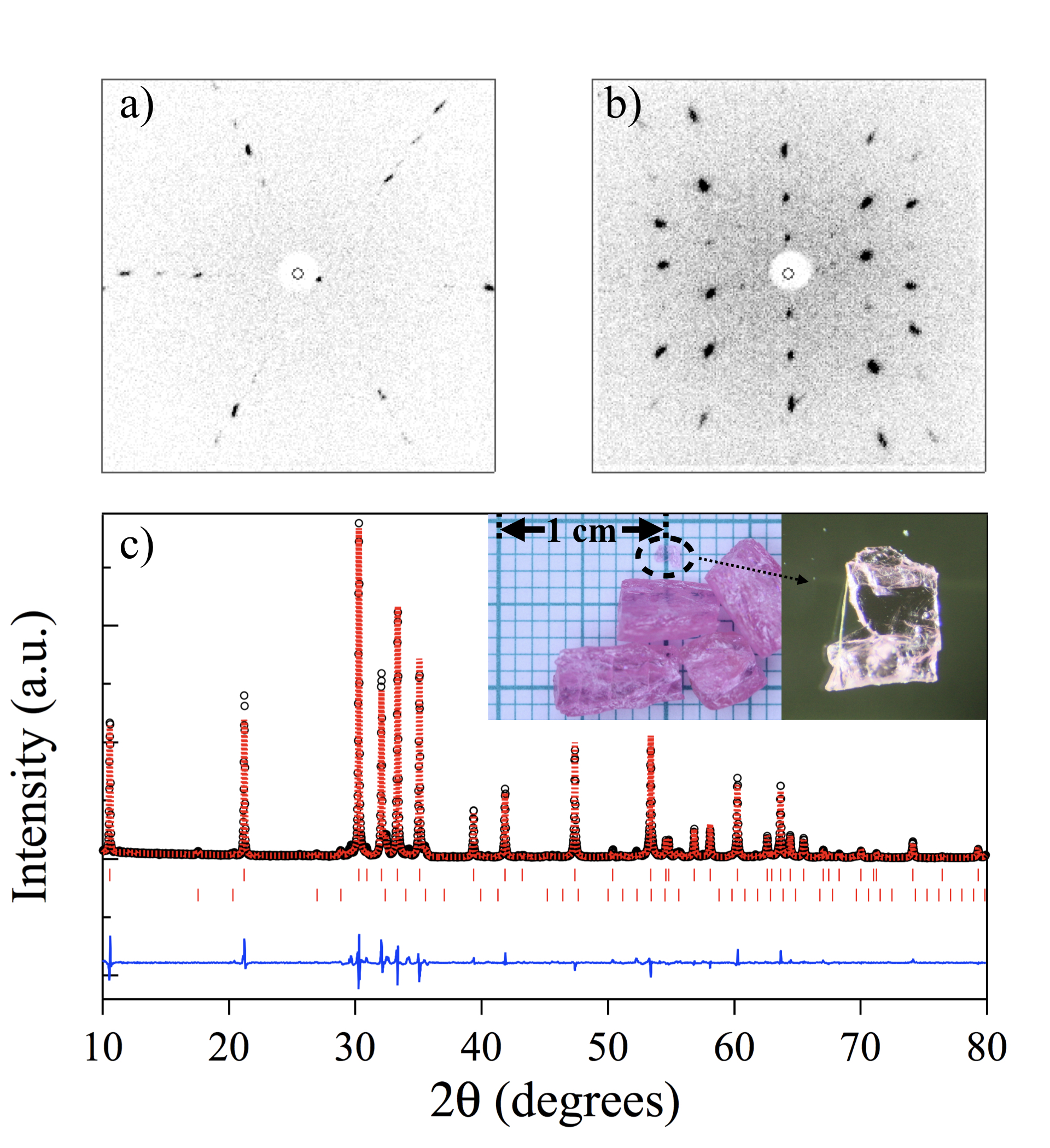}
\caption{Structural chracterization of ErMgGaO$_4$ by Laue diffraction and powder X-ray diffraction: a)~\& b) the Laue pattern in orientation of (001) and (100), c) Rietveld refinement of powder X-ray diffraction data. (Insert shows a picture of crystals and an extracted transparent small crystal)}
\label{RietveldRefinement}
\end{figure}
Crystal growth was carried out in a two-mirror optical floating zone image furnace (NEC systems). The optimal growing conditions were found from multiple attempts to have oxygen gas at 2 atmosphere overpressure and at a growth rate of 0.5~mm/hr. 
No clear large single crystal was found even after 8 cm growth, instead, transparent, relatively large multi-grain crystals (approximately 1cm in length) were obtained. Those grains were found to be misaligned in the ab-plane, but aligned with their c axis (which was usually perpendicular to the growth direction); this pattern of crystal formation of the crystal likely reflects the huge lattice parameter difference $c \gg a$. These multi-grain crystals are shown in Fig.~\ref{RietveldRefinement}: we also extracted single crystals from them. In Fig.~\ref{RietveldRefinement}, we present the characterization of a separated small crystal utilizing Laue X-ray diffraction. We also ground up a small crystal for powder X-ray diffraction. Structural Rietveld refinements were carried out using Fullprof software package. As shown in Fig.~\ref{RietveldRefinement}c, refinements to our X-ray diffraction data yield a good fit within two phases: ErMgGaO$_4$ (R{$\overline3$}m) and Er$_3$Ga$_5$O$_{12}$ (Ia{$\overline3$}d) with $\chi^2 = 3.96$ and weight fractions as 95.11$\%$ and 4.89$\%$, respectively. Our Laue diffraction measurements confirm the R{$\overline3$}m structure as seen in Fig.~\ref{RietveldRefinement} a$\&$b with the crystal oriented in the a-c plane.

\begin{figure}
\includegraphics[width=\columnwidth]{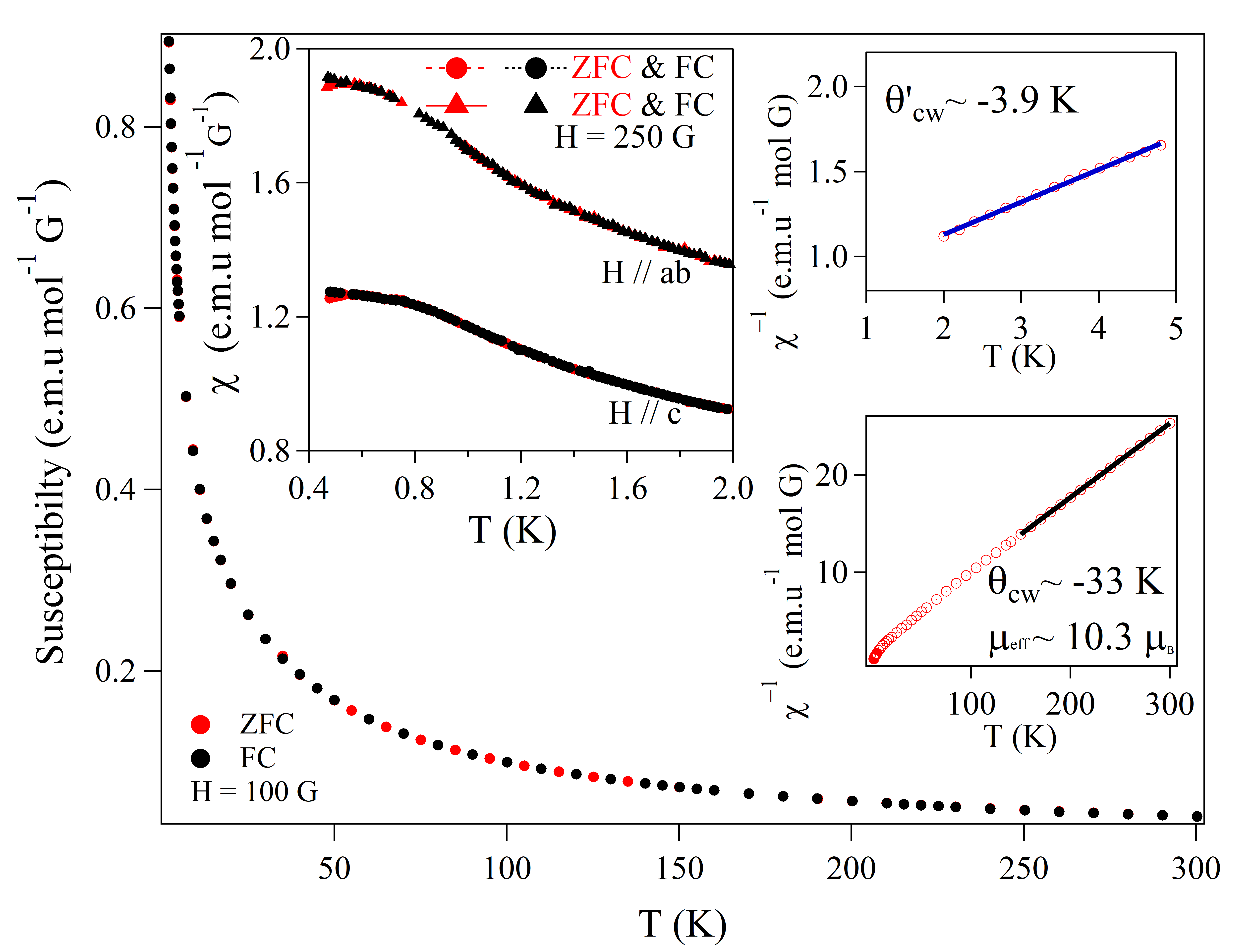}
\caption{Main panel: temperature dependent DC susceptibility from 2~K to 300~K at 100~G. Left insert: temperature dependent DC susceptibility below 2~K with applied external field along selected orientation. Right insert: the inverse susceptibility data (red dot) at temperatures between 2 to 5~K and 150 to 300~K with the Cure-Weiss fit superimposed on the data as solid lines.}
\label{susceptibility}
\end{figure}

We performed magnetic susceptibility measurements on a small crystal from 0.48~K to 300~K with a Quantum Design MPMS XL-3 equipped with an iQuantum He$^3$ Insert for measurements below 2~K. We also measured the specific heat between 0.1~K and 4~K using a Quantum Design PPMS with a dilution refrigerator insert.

As shown in Fig.~\ref{susceptibility}, the magnetic susceptibility at high temperature (150-300~K) was well described by a Curie Weiss Law with the resulting Weiss temperature $\theta_{CW}$ = -33~K and an effective Er moment of 10.3(1)~$\mu_{B}$; the latter corresponds fairly well to the expected value of 9.59 $\mu_{B}$ for an isolated Er$^{3+}$ ion. The Curie-Weiss fit to the magnetic susceptibility in the low temperature regime (2-5~K) results in a Weiss temperature $\theta^{'}_{CW}$ = -3.9~K. Further measurements below 2~K indicates the presence of magnetic anisotropy and no evidence of any magnetic transitions down to 0.5~K in this compound. The high temperature results as well as magnetic anisotropy properties (above 2~K) are consistent with previous measurements \cite{Cevallos2}. Furthermore, there is no splitting of the zero-field-cooled and field-cooled susceptibilities other than a very small signature from Er$_3$Ga$_5$O$_{12}$ \cite{19_Cai}, which rules out the presence of a glassy spin freezing transition. The dominant antiferromagnetic exchange interaction in ErMgGaO$_4$ with no sign of magnetic transitions reveals that magnetic frustration is significant, which is similar to YbMgGaO$_4$ \cite{Li2} and TmMgGaO$_4$ \cite{Cevallos1}.

One important feature of the ground state properties of ErMgGaO$_4$ is its crystal electric field (CEF) induced magnetic spin anisotropy, where $Er^{3+}$ has trigonal local symmetry with the point group $D_{3d}$. Spin anisotropy studies in YbMgGaO$_4$ have shown that a point charge calculation under a single ion approximation well captures its major properties, which were later found to be in agreement with their inelastic neutron scattering measurements \cite{Yuesheng}. We performed a similar point charge calculation under the single ion approximation to examine the ground state magnetism of ErMgGaO$_4$. According to Hund's rules, the total angular momentum of the $Er^{3+}$ ion is $J = 15/2$ and the $(2J+1)$-fold degeneracy is lifted by CEFs due largely to the presence of the neighbouring $O^{2-}$ions into 8 Kramers doublets. Neglecting the potential effect of crystallographic distortion from the Ga/Mg site mixing, the CEF Hamiltonian is written as 

\begin{equation}
\label{eq1}
{\hat{\cal H}}_{D_{3d}}^{CEF} = B_2^{0}\hat{O}_2^{0} + \sum_{i=0,3} B_4^{i}\hat{O}_4^{i} +\sum_{i=0,3,6} B_6^{i}\hat{O}_6^{i},
\end{equation}
 where the $B_n^i$ are rare-earth dependent coefficients and $\hat{O}_n^i$ are Stevens operators which are combinations of total angular momentum operators \cite{Hutching}. 
 
In Table.~\ref{CEF_coefficient}, we present our results obtained from this point charge calculation. This calculation yielded a well separated ground state and the ground state doublet is comprised primarily of $m_J = \pm15/2$. The corresponding anisotropic $g$-tensor components are given by $g_\parallel = 2g_J|\langle \phi_0^{\pm}|J_z|\phi_0^{\pm}\rangle| = 17.91$ and $g_\perp = g_J|\langle \phi_0^{\pm}|J_{\pm}|\phi_0^{\mp}\rangle| = 0$, where $z$ corresponds to the local c axis which implies a strong, effectively, Ising-like behaviour of the $Er^{3+}$ moment along the c axis. We caution that the point charge calculation ignores the effect beyond nearest six oxygen neighbours and does not account the effects of the distorted electrostatic potential due to the $Ga/Mg$ site mixing. This distortion in YbMgGaO$_4$ was found to only slightly broaden the CEF excitations and present a distribution of g-tensors but without dramatically changing its magnetic properties \cite{Yuesheng}. Our finding of Ising-like moments is quite different from YbMgGaO$_4$, where a Heisenberg-like spin anisotropy was established \cite{Yuesheng, Paddison}.
 
 \begin{table}
\caption{The CEF parameters and eigenvalues for ErMgGaO$_4$ from point charge calculation. The ground state eigenvector are given in terms of the $m_J$ basis with $J = 15/2$ for Er. }
\label{CEF_coefficient}
\begin{ruledtabular}
\begin{tabular}{lllllll} 
$B_2^0 (meV)$ & $B_4^0$ & $B_4^3$ & $B_6^0$ & $B_6^3$ & $B_6^6$ \\
-0.3440 & -5.3176e-4 & -0.0122 & 2.4242e-6 & 1.3236e-5 & 7.043e-20
\end{tabular}
\begin{tabular}{lllllllll} 
$Calcu$(meV) & 0& 19.06& 36.99 & 48.63 & 52.91 & 61.71 & 62.69 & 64.37 \\
\hline \\[1pt]
\multicolumn{9}{c}{$\begin{aligned} 
\phi_0^{\pm} = \mp 0.9937\mid\mp{\frac{15}{2}}\rangle + 0.1111\mid\mp{\frac{9}{2}}\rangle \mp 0.0128\mid\mp{\frac{3}{2}}\rangle;
\end{aligned}$} \\ 
\\[1pt]
\multicolumn{9}{c}{$[g_\parallel, g_\perp] = [17.91, 0]$} 

\end{tabular}
\end{ruledtabular}
\end{table}

\begin{figure}
\includegraphics[width=\columnwidth]{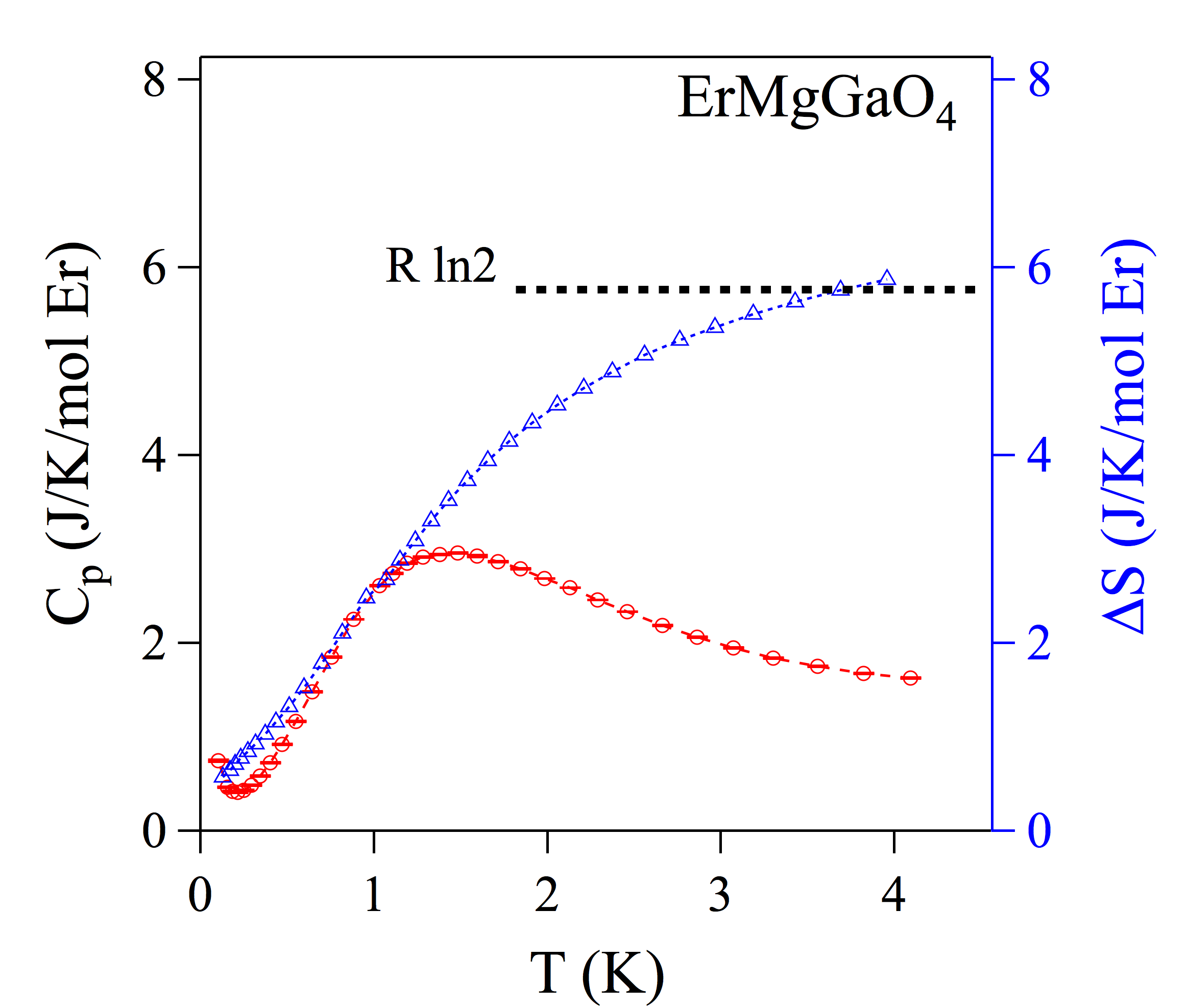}
\caption{Red circle points shows the specific heat versus temperature for ErMgGaO$_4$. The blue triangular shows the temperature dependent integral of $C_p/T$.}
\label{Fig: specific heat}
\end{figure}

We next turn to the low-temperature collective magnetic properties of ErMgGaO$_4$. We performed a specific heat measurement in zero field to look for signs of magnetic order, low-lying magnetic excitations, and residual entropy. In Fig.~\ref{Fig: specific heat}, we display our specific heat data for ErMgGaO$_4$ in the low-temperature regime. The upturn at the lowest temperatures likely arises from a $^{167}$Er nuclear Schottky contribution. With increasing temperature, only a broad hump is seen with its maximum at around 1.5 K. No clearly $\lambda$ anomaly was observed which would be expected for any magnetic order. We also plot the entropy recovery when warming through the hump in Fig~\ref{Fig: specific heat}, which saturates near $Rln(2)$ by 4~K. This finding is consistent with the well-isolated CEF doublet ground state as we expect on the basis of our previous CEF anisotropy calculation and indicates that there is essentially no residual entropy remaining as temperature approaching zero. This result also implies there will not be any further magnetic transitions down to zero temperature and that similar to YbMgGaO$_4$, ErMgGaO$_4$ is a quantum spin liquid candidate.

To directly rule out any possible magnetic ordering, we turned to muon spin rotation and relaxation ($\mu$SR), the most sensitive technique in detecting such weak ordering. In addition, $\mu$SR is also capable to characterize the low energy spin dynamics, which is also crucial in examining a quantum spin liquid state. We performed $\mu$SR measurements on the M15 and M20 beam-lines at the TRIUMF laboratory in Vancouver, Canada. Crystals of ErMgGaO$_4$ (aligned within 5 degrees along c-axis) were mounted on the M20 beam-line in a low background apparatus utilizing a He$^4$ cryostat, and later mounted onto an Ag plate and covered in thin Ag foil for the measurements in a dilution refrigerator on the M15 beamline. We performed measurements in both zero applied field (ZF) and with a magnetic field applied along the incident muon spin direction (LF). All the $\mu$SR data were fit by the open source $\mu$SRfit software package \cite{A. Suter}.

\begin{figure}
\includegraphics[width=\columnwidth]{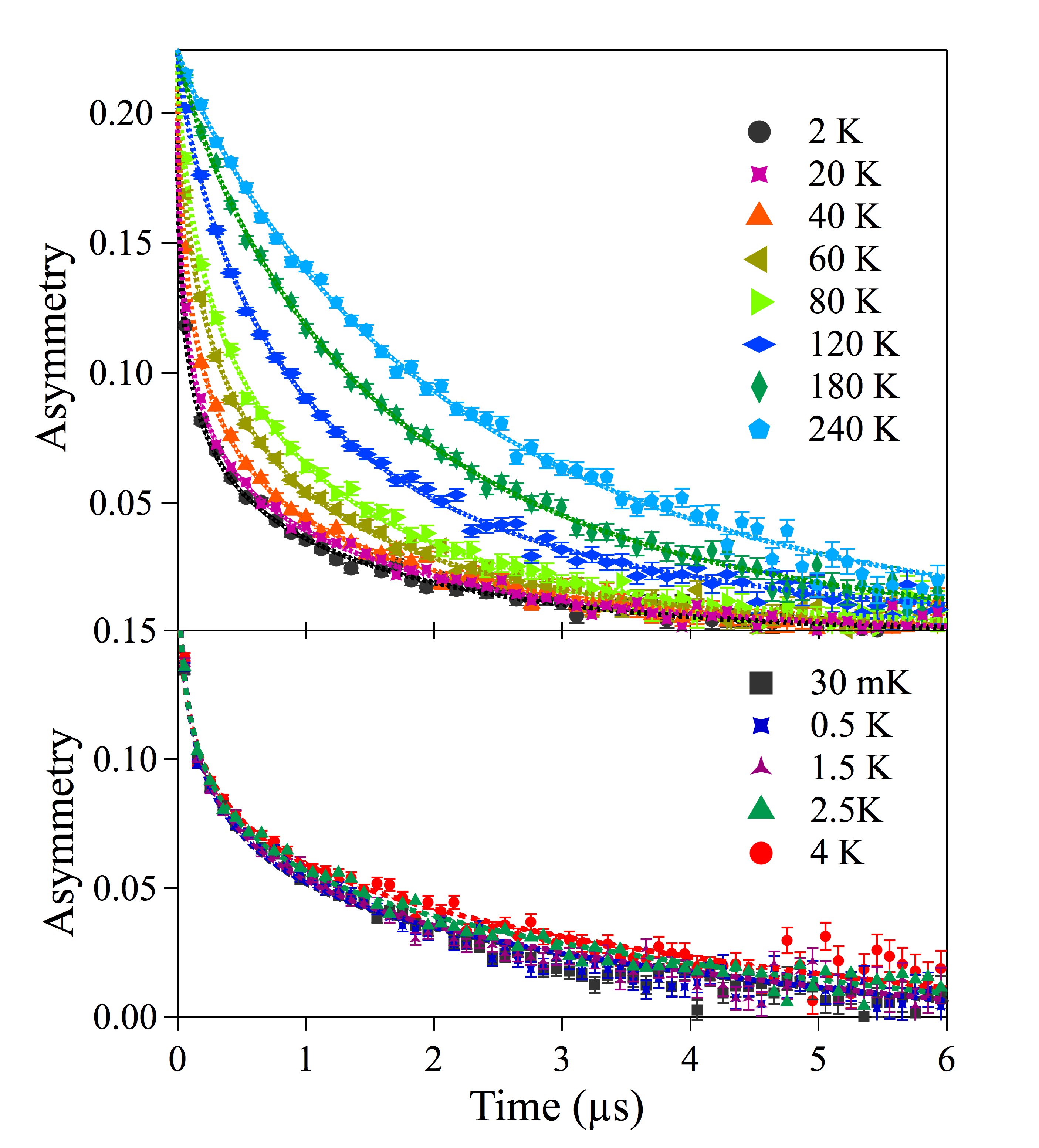}
\caption{Zero field $\mu$SR spectra of ErMgGaO$_4$ measured at temperature range 25~mK to 300~K. Coloured dots are experiment data, and corresponding dashed lines are the fitting results as described in text}
\label{ZF_spectra}
\end{figure}

\begin{figure}
\includegraphics[width=\columnwidth]{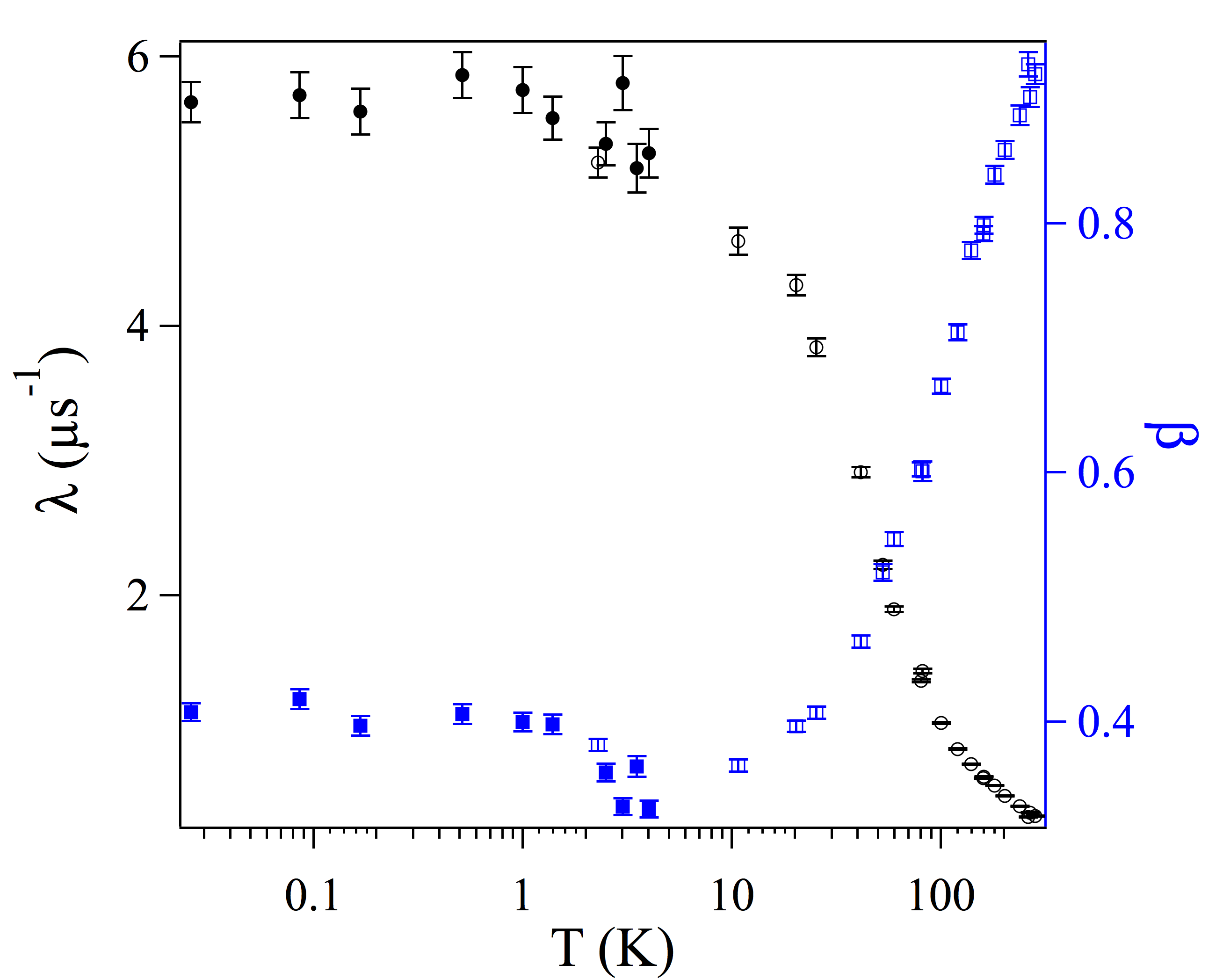}
\caption{Temperature dependence of zero field $\mu$SR relaxation rate in ErMgGaO$_4$. Open symbols are measurements obtained from He$^4$ cryostat ($>$~2 ~K), and closed symbols are measurements obtained from dilution refrigerator ($< $~ 5~K) }
\label{Fig:relaxation_rate}
\end{figure}

In Fig.~\ref{ZF_spectra}, we show ZF$-\mu$SR data for ErMgGaO$_4$ between 25 mK and 300 K. The relaxation rate keeps increasing as the temperature deceases and the data exhibit no sign of oscillations down to 25 mK. The absence of spontaneous muon precession indicates there is no transition to long-range magnetic order in ErMgGaO$_4$, which is consistent with the magnetic susceptibility and specific heat results discussed above. The ZF$-\mu$SR asymmetry spectra relax to the same zero baseline without a recovery of 1/3 tail which rules out the possibility of a spin glass state, which would have 1/3 of muon polarization parallel to the random local field \cite{Yaouanc_uSR}. At all temperature, the ZF$-\mu$SR spectra were well fit to a stretched exponential: $$P(t) = A_{total} {e^{-(\lambda t)^\beta}}$$ where the asymmetry $A_{total}$ was independent of temperature.

The temperature dependence of the relaxation rate $\lambda$ as well as the exponent $\beta$ are shown in Fig.~\ref{Fig:relaxation_rate}. Above $\sim$ 2 K, a sharply decreasing relaxation rate was observed with temperature increasing up to 300K. This behaviour is likely due to the Orbach process \cite{Orbach}: the thermal excitation of the crystal field levels. The relaxation rate appears to saturate and becomes temperature independent from about 2~K down to 25~mK with no sign of magnetic ordering in ErMgGaO$_4$. The value of relaxation rate below 2~K, $\lambda \sim 6~\mu s^{-1}$, is considerably larger than the corresponding value found in YbMgGaO$_4$ (0.3~$\mu s^{-1}$) \cite{Yuesheng_uSR}, this difference is likely due to the larger moment from $Er^{3+}$ compared to $Yb^{3+}$.

\begin{figure}[h]
\includegraphics[width=\columnwidth]{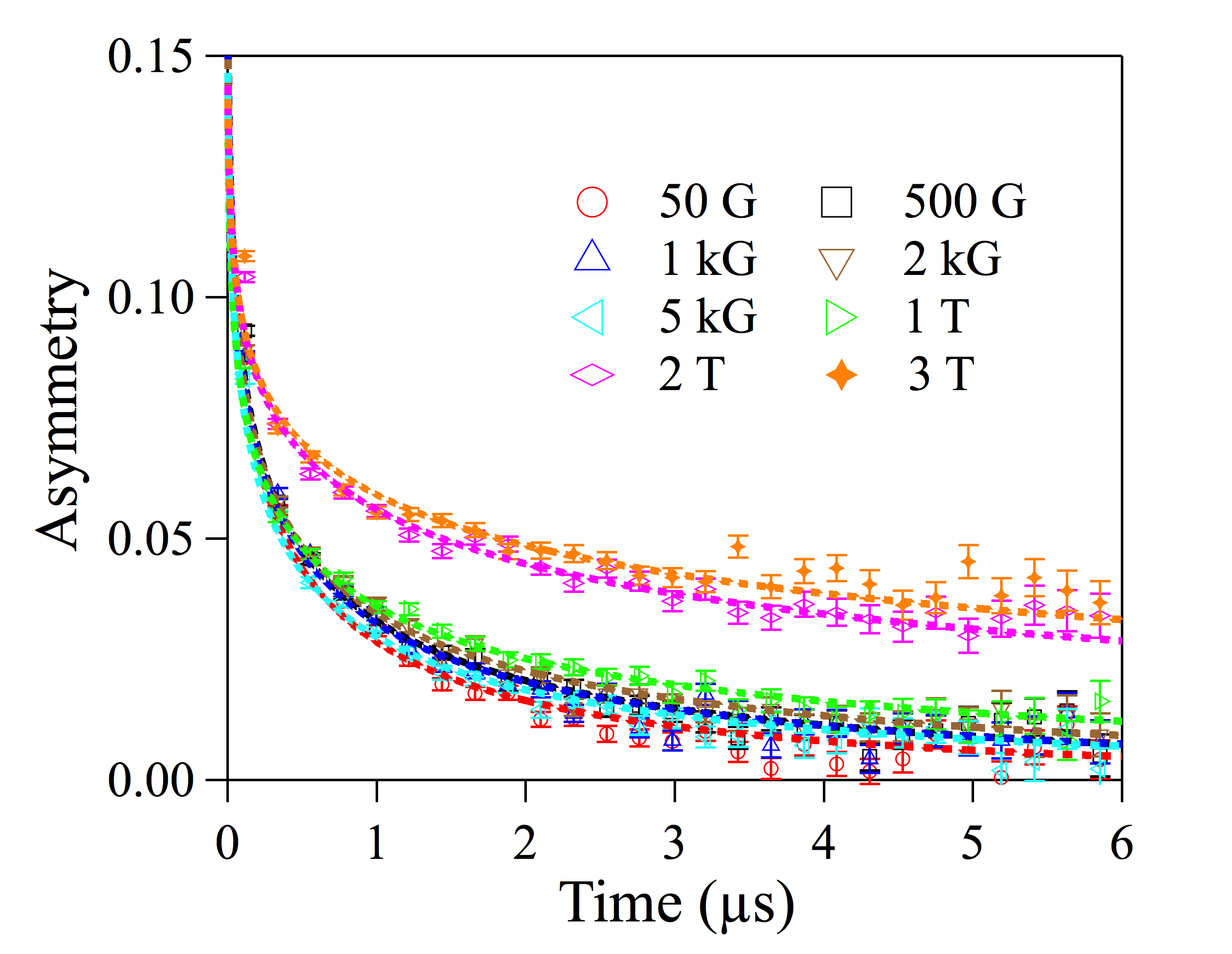}
\caption{$\mu$SR measurements on ErMgGaO$_4$ at 25~mK in selected longitudinal field up to 3~T.}
\label{Fig:LF}
\end{figure}
We also performed longitudinal field (LF) $\mu$SR measurements to test for the presence of spin fluctuations. In a LF-$\mu$SR setup, the external field is applied in the direction of the initial muon spin polarization. In the case of a quasi-static internal field distribution, a static relaxation signal will be nearly fully decoupled by an applied field that is a few times larger than the field corresponding to the ZF relaxation rate. However, if the relaxation of the ZF-$\mu$SR signal comes from fluctuating fields, the signal will not be decoupled by an applied field of this magnitude. In this case, the signal will only slowly decouple, and relaxation will continue to be apparent up to relatively large applied longitudinal fields \cite{Uemura3}. If the fluctuation rate is independent of the applied field, then the relaxation rate will decrease with applied field according to the Redfield form \cite{redfield}, where a field corresponding roughly to the fluctuation rate will be needed to greatly decrease the relaxation. In Fig.~\ref{Fig:LF}, we present our LF scans at 25~mK, the relaxation rate at this temperature is barely decoupled even up to 1~T (much bigger than the local field $\lambda_0/\gamma_{\mu} \sim$ 66~G, we would infer from our ZF spectra if the origin of the ZF relaxation were quasi-static internal fields). This means the spins associated with the $Er^{3+}$ ions remain in a dynamically fluctuating state down to our lowest temperature 25~mK, consistent with a quantum spin liquid state. Similar spin dynamics were found in YbMgGaO$_4$ with estimated local fields $\sim$ ~ 0.09~mT, where only small decoupling was observed with applied LF field up to 0.18~T \cite{Yuesheng_uSR}.

In conclusion, we have grown single crystals and have performed a detailed study of magnetism and spin dynamics in ErMgGaO$_4$. Our ZF-$\mu$SR, specific heat and magnetic susceptibility measurements reveals no presence of static internal magnetic fields or magnetic transition. LF-$\mu$SR measurements detect the existence of persistent spin fluctuations down to our lowest temperature $\sim$ 25~mK. Our observations provide evidence of a quantum spin liquid state in the triangular antiferromagnet ErMgGaO$_4$. Point charge calculations suggest that the Er moments are predominately Ising-like, in contrast to the Heisenberg spins found in YbMgGaO$_4$; however inelastic neutron scattering measurements to experimentally determine the CEF parameters would be needed to confirm this.

We greatly appreciate the support of personnel at TRIUMF during the $\mu$SR measurements. Work at McMaster was supported by the Natural Sciences and Engineering Research of Council of Canada.

\end{document}